\documentclass[aps,twocolumn]{revtex4}
\usepackage{dcolumn}
\usepackage{epsfig}

\begin{document}

\title{Large difference in the elastic properties of fcc and hcp
  hard-sphere crystals}
\author{Sander Pronk}
\author{Daan Frenkel}
\affiliation{
FOM Institute for Atomic and Molecular Physics\\ 
Kruislaan 407\\
1098 SJ Amsterdam, 
The Netherlands}

\begin{abstract}
  We report a numerical calculation of the elastic constants of the
  fcc and hcp crystal phases of monodisperse hard-sphere colloids.
  Surprisingly, some of these elastic constants are very different (up
  to 20\%), even though the free energy, pressure and bulk
  compressibility of the two crystal structures are very nearly equal.
  As a consequence, a moderate deformation of a hard-sphere crystal
  may make the hcp phase more stable than the fcc phase. This finding
  has implications for the design of patterned templates to grow
  colloidal hcp crystals.  We also find that, below close packing,
  there is a small, but significant, difference between the distances
  between hexagonal layers ($c/a$ ratios) of fcc and hcp crystals.
\end{abstract}

\maketitle

\begin{figure}
  \includegraphics[angle=0,width=8.5cm]{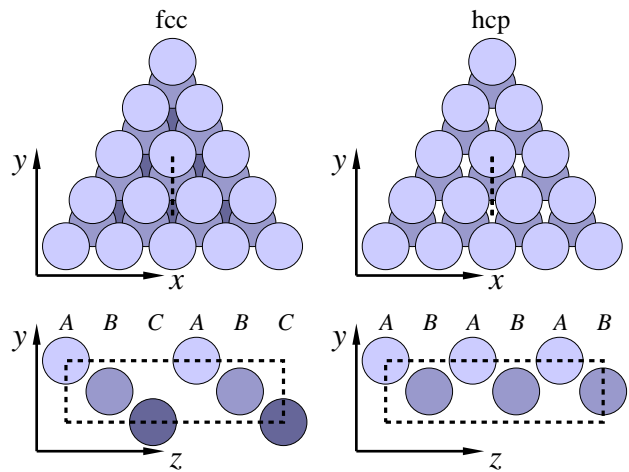}
  \caption{Sketch of the structures of the regular close-packed fcc 
    crystal (left) and hcp crystal (right).  Inequivalent close-packed
    layers are labeled with the letters A,B and, in the case of fcc,
    C.  An fcc crystal has $ABCABC\ldots$ stacking and an hcp crystal
    has $ABABAB\ldots$ stacking. The $c/a$ ratio is the distance
    between two close-packed layers divided by the distance between
    neighboring particles in a close-packed layer. The figures show
    the definitions of the $x$,$y$ and $z$ directions referred to in
    the text.  }
  \label{coordframe}
\end{figure}

The simplest regular close-packed structures of hard, spherical
particles are the face-centered cubic (fcc) and hexagonal close-packed
(hcp) structures (see Fig.~\ref{coordframe}). Close to melting,
the Helmholtz free energies of these two crystal structures
differ by less than
0.05\%~\cite{BolhuisNatureFccHcp,BruceWildingLatticeSwitch,
  PronkFrenkelStacking}. As a consequence, hard-sphere colloids (the
experimental realization of elastic hard spheres) rarely crystallize
directly into the more stable fcc structure. Rather, crystallization
initially results in the formation of a randomly stacked
crystal~\cite{ZhuChaikinMicrograv,PetukhovBraggSpots}. The latter then
slowly transforms to the stable fcc
structure~\cite{ChengMicrogravGrowing,KegelDhontAging,PoonStacking,
  PronkFrenkelStacking}. However, pure \emph{hcp} crystals have
recently been grown by colloidal epitaxy on patterned templates
\cite{JacobHCP}.  At a given density, not only the free energies, but
also the pressures and compressibilities of the fcc and hcp phases are
very similar. One might therefore be tempted to suppose that these two
crystal phases are similar in {\em all} their thermodynamic
properties. Surprisingly, this is not the case. In this Letter we
present calculations of the elastic constants of fcc and hcp
hard-sphere crystals. We show that some of these elastic constants may
differ by as much as 20\%. As a consequence, a moderate
deformation of the hard-sphere crystal may change the relative
stability of the two crystal phases.

A homogeneous deformation of a solid can be described by the
transformation matrix $\alpha_{ij}$ that relates the cartesian
coordinates $x_j$ of a point in the undeformed solid, to the
coordinates $x'_i$ in the deformed solid: $x'_i=\alpha_{ij}x_j$, where
summation of repeated indices is implied.  The (isothermal) elastic
constants of a crystal are most easily defined in terms of an
expansion of the Helmholtz free energy $F(N,V,T)$ in powers of the
Lagrangian strains $\eta_{ij}$~\cite{WallaceElastic}:
\begin{eqnarray}
\label{eq:elastic1}
  F(\eta_{ij})/V & = & F(\mathbf{0})/V + T_{ij}(\mathbf{0})\eta_{ij} +
  \frac{1}{2} C_{ijkl} \eta_{ij}\eta_{kl} \nonumber\\
  & & +  \frac{1}{6} C_{ijklmn} \eta_{ij}\eta_{kl}\eta_{mn} + \ldots
  \label{ElasticFreeEnergy}
\end{eqnarray}
The Lagrangian strain parameters $\eta_{ij}$ are related to the
deformation matrix through $\eta_{ij}\equiv
\frac{1}{2}(\alpha_{ki}\alpha_{kj}-\delta_{ij})$. In
Eq.~\ref{eq:elastic1}, the coefficients $T_{ij}(\mathbf{0})$ are
simply the components of the stress tensor at zero deformation,
$C_{ijkl}$ are the second-order elastic constants, $C_{ijklmn}$ are
the third-order elastic constants, and so on. For a system under
hydrostatic pressure $P$, the components of the stress tensor are
$T_{ij} =-\delta_{ij}P$.

The fcc lattice has only 3 independent elastic constants
\cite{WallaceElastic} ($C_{1111}\equiv C_{11}$, $C_{1122}\equiv
C_{12}$ and $C_{2323}\equiv C_{44}$ in the coordinate frame of the
cubic unit cell).  In what follows, we use this Voigt notation
($C_{ij}$) to denote the second-order elastic constants.

In order to
compare the elastic constants of the fcc and hcp crystals, we used the
coordinate system shown in Fig.~\ref{coordframe}, with the $x$ and $y$
directions in the hexagonal planes and the $z$ direction perpendicular
to these planes. For hcp (with hexagonal symmetry), there are six
distinct elastic constants, five of which are
independent\cite{WallaceElastic}.  To make a term-by-term comparison
of the fcc and hcp elastic constant, it is convenient to ignore the
full symmetry of the fcc crystal, and only use the fact that the
crystal also has a lower rhombohedral symmetry.  If the symmetry were
really rhombohedral, the fcc crystal would have six independent
elastic constants. But, if we take the full fcc symmetry into account,
only three are linearly independent; the usual fcc elastic constants
can be expressed as linear combinations of the rhombohedral elastic
constants $C'_{ij}$: $C_{11} = 4 C'_{11} - 3 C'_{33}$, $C_{12}=
C'_{33} + C'_{12} - C'_{11}$ and $C_{44} = C'_{33}
-\frac{1}{2}(C'_{11} + C'_{12})$.

We computed the elastic constants by calculating the stress response
to a small applied strain, using molecular dynamics
simulations~\cite{FrenkelLaddElastic}. At zero deformation, the stress
response of a system with isotropic pressure $P$ is given by a
generalization of Hooke's law:
\begin{equation}
  \frac{\partial T_{ij}}{\partial \alpha_{kl}} = \left(
    \delta_{ij}\delta_{kl} -\delta_{il}\delta_{jk} - \delta_{jl}\delta_{ik}
     \right)P + C_{ijkl}
     \label{hooke}
\end{equation}
For the MD simulations, we used the event-based algorithm described by
Rapaport\cite{RapaportMD}. The pressure tensor is calculated as the
time average of the dyadic product of the collisional momentum
exchange vector and the particle separation vector for each
two-particle collision\cite{AllenTildesley}.

We performed simulations for a range of amplitudes of each type of
deformation.  The second-order elastic constants were deduced from the
linear part of the stress-strain relation.  In principle, all elastic
constants can also be calculated in a single simulation using
fluctuation methods
\cite{ParrinelloFluct,SprikElastic,WojciechowskiOrientPent}.  However,
these methods suffer from slow convergence \cite{SprikElastic}. We
found the stress-strain method to be the most efficient.

For some deformations, we also computed the third-order elastic
constants from the second derivative of the stress tensor with respect
to deformation:
\begin{eqnarray}
  \label{SecondDeriv}
  \frac{\partial^2 T_{ij}}{\partial \alpha_{rs} \partial \alpha_{tu}} & = &
  2\delta_{tu}\delta_{rs}T_{ij}
  + (\delta_{it}\delta_{jr}+\delta_{ir}\delta_{jt})T_{su}\nonumber\\
  & &
  - \delta_{ut}(\delta_{ir}T_{js} +\delta_{jr}T_{is})\nonumber\\
  & &
  - \delta_{sr}(\delta_{jt}T_{iu} +\delta_{it}T_{uj})\nonumber\\
  & &- \delta_{ut}C_{ijrs} 
  + \delta_{it}C_{ujrs}
  + \delta_{jt}C_{iurs}  \nonumber\\
  & &- \delta_{sr}C_{ijtu} + 
  \delta_{ir}C_{sjtu}
  + \delta_{jr}C_{istu}  \nonumber\\
  & & + \delta_{rt}C_{ijsu} + C_{ijrstu}
\end{eqnarray}
The third-order elastic constants $C_{ijrstu}$ appear in the last
term.

\begin{table*}
  \begin{tabular}{l l l|l l l l l l}
    \hline
    \hline
    $\phi$ & $N$ & & $C'_{11}$ & $C'_{12}$ & $C'_{13}$ 
            &$C'_{14}$ &$C'_{33}$ & $C'_{44}$\\
    \hline
    $0.543$ & 13292 & fcc & 90.51(6) & 13.56(7) & 7.51(7) 
    & -8.77(4) & 96.7(1) & 32.22(6) \\
            &       & hcp & 87.39(8) & 15.95(7) & 7.7(1) 
            & 0 & 96.56(9) & 33.79(4) \\
            &       & \em{hcp} & \em{87.0(1)} & \em{15.82(9)} 
            & \em{7.83(8)} & \em{0} &
              \em{97.1(1)} & \em{33.90(5)} \\
    $0.543$ & 216 & fcc & 90.50(8) & 13.8(1) & 7.57(8) 
    & -8.75(6) & 97.0(1) & 32.4(1)  \\
            &     & hcp & 87.39(7) & 16.6(1) & 7.56(9) 
            & 0 & 96.67(9) & 35.0(1)  \\
    \hline
    $0.550$ & 216 & fcc & 99.41(9) & 15.2(1) & 8.4(1) 
    & -9.65(4) & 106.16(8) & 35.76(4)\\
            &     & hcp & 95.88(6) & 17.9(1) & 8.6(1) 
            & 0 & 106.1(1) & 37.38(7) \\
    \hline
    $0.576$ & 13292& fcc & 146.42(8) & 21.86(7) & 12.1(1) 
    & -13.82(6) & 156.1(1) & 52.33(5) \\
            &      & hcp & 142.1(1) & 25.64(7) & 12.36(8) 
            & 0 & 155.78(9) & 54.56(4)\\
    $0.576$ & 216 & fcc & 146.1(1) & 21.8(2) & 12.1(1) 
    & -14.3(1) & 156.3(3) & 52.8(4) \\
            &     & hcp & 141.8(1) & 25.8(1) & 12.44(9) 
            & 0 & 156.1(4) & 54.9(1) \\
    \hline
    $0.628$ & 216 & fcc & 366.4(6) & 51.6(4) & 26.4(5) 
    & -35.4(1) & 392(1) & 133.7(2) \\
            &     & hcp & 356.9(4) & 60.3(6) & 27.3(3) 
            & 0 & 390(1) & 138.2(1) \\
    \hline
    $0.681$ & 216 & fcc & 1463(3) & 189(2) & 89(2) 
    & -145(2) & 1563(3) & 535(2) \\
            &     & hcp & 1423(3) & 223(3) & 97(1) 
            & 0 & 1559(2) & 557(2)  \\
    \hline
    $0.733$ & 216 & fcc & $1.10(1)\cdot10^5$ & $1.28(1)\cdot10^4$ &
    $6.1(2)\cdot10^3$ & $-1.05(3)\cdot10^4$ & $1.17(2)\cdot10^5$ &
    $4.05(4)\cdot10^4$  \\
            &     & hcp & $1.08(1)\cdot10^5$ & $1.52(1)\cdot10^4$ &
            $5.4(1)\cdot10^3$ & 0 & $1.17(1)\cdot10^5$ &
            $4.08(1)\cdot10^4$ \\
 \end{tabular}
 \caption{
   Second-order elastic constants of fcc and hcp hard-sphere crystals at 
   densities between the melting point and close packing. The values for 
   the hcp structure with $c/a=\sqrt{8/3}$ are shown in upright font. 
   The (almost identical) results for a fully relaxed $c/a$ ratio: 
   $c/a=\sqrt{8/3}(1-7.5\cdot10^{-4})$ at $\phi=0.543$, 
   are shown in italics. 
   The simulation equilibration time was $1\cdot 10^4$ collisions per particle.
   Data were collected during typically $2\cdot 10^6$ collisions per
   particle for the $216$ particle system, and $6\cdot 10^4$ collisions
   per particle for the $13292$ particle system. For each deformation,
   $8$ simulations were done at different strain amplitudes to check linearity 
   of the stress response.  The calculations of the stress-strain curve 
   for each type of deformation involved simulations totaling several billion
   collisions $6.4 \cdot 10^9$ collisions
   (one week on an Athlon 1600+ CPU). 
   }
 \label{fulltable}
\end{table*}

The simulations were performed on systems with $6\times6\times6=216$,
$12\times12\times12=1728$ and $24\times24\times24=13824$ particles.
The maximum applied deformation at lower densities was
$4\cdot10^{-3}$; higher densities required even smaller deformations
to keep the stress response linear. 
The measured elastic constants
between the melting point (packing fraction
$\phi=0.54329$~\cite{FrenkelSmit}) and close packing are given in
table \ref{fulltable}.

\begin{figure}
  \includegraphics[angle=0,width=8.0cm]{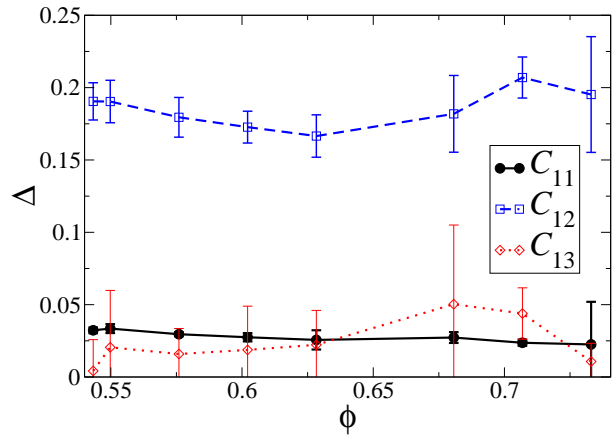}
  \caption{
    Relative difference $\Delta \equiv |C'^{\mathrm{fcc}}_{ab}-
    C'^{\mathrm{hcp}}_{ab}|/C'^{\mathrm{fcc}}_{ab}$ between fcc and
    hcp $C'_{11}$, $C'_{12}$ and $C'_{13}$ elastic constants as a
    function of packing fraction $\phi$. The results shown were
    obtained in simulations of a system of 216 particles, with
    $c/a=\sqrt{8/3}$. The curves only serve as guides to the eye.}
  \label{highdens}
\end{figure}

At all densities, the values of the fcc and hcp elastic constants
differ significantly (see Fig.~\ref{highdens}). The relative
differences between the elastic constants appear to remain
approximately constant over the entire density range. The largest
difference between fcc and hcp (up to 20\%) was found for $C'_{12}$.
Yet, the compressibilities of the two phases are identical to within
the measurement error. For instance, at melting:
$K^T_{fcc}=0.02422(5)$ vs. $K^T_{hcp}=0.02424(5)$ (for 1728
particles). We computed these compressibilities in two ways: (a) from
the appropriate linear combination of elastic constants and (b)
directly from the equation of state~\cite{SpeedyCrystal}.  The results
are the same, to within the statistical error. At the same density,
the pressures of the fcc and hcp phases are also very similar:
$P_{fcc}=11.568(1)$ and $P_{hcp}=11.571(1)$.  Finally, the free
energies differ only by about $1.12(4)\;10^{-3} k_BT$ per particle
\cite{BolhuisNatureFccHcp,BruceWildingLatticeSwitch,PronkFrenkelStacking}.

\begin{figure}
  \includegraphics[angle=0,width=8.0cm]{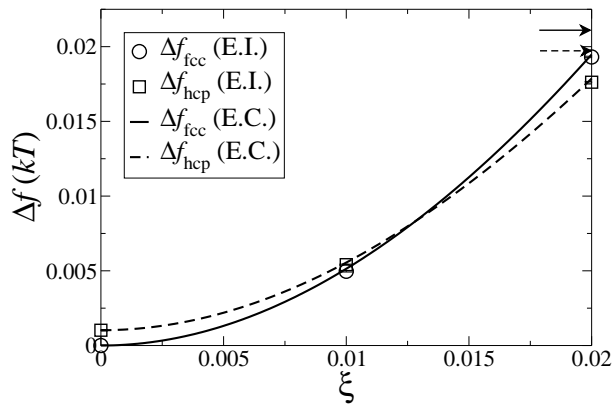}
  \caption{
    Variation of the fcc and hcp free energy with deformation $\xi$
    (see Eq.~\ref{DeformationMatrix}). The symbols indicate the
    results of Einstein free energy calculations (see text).  The
    continuous curves were computed on basis of the calculated second
    and third order elastic constants. The error bars of the Einstein
    free-energy calculations are about one sixth the size of the
    symbols.  The two horizontal arrows show the predictions for
    $\Delta f$ obtained by neglecting the third-order elastic
    constants.}
  \label{figdf}
\end{figure}

\begin{table}
    \begin{tabular}{l|l l l l l l l}
      \hline
      \hline
      & fcc & hcp \\
      \hline
      $C'_{111}$ & $-2.0(1)\cdot10^{3}$ & $-2.1(1)\cdot10^{3}$ \\
      $C'_{112}$ & $-7.3(9)\cdot10^{1}$ & $-7.9(9)\cdot10^{1}$ \\
      $C'_{122}$ & $-3.2(9)\cdot10^{2}$ & $-4.2(8)\cdot10^{2}$ \\
      $C'_{222}$ & $-1.71(8)\cdot10^{3}$ & $-1.71(8)\cdot10^{3}$ \\
      \hline
    \end{tabular}
  \caption{Values for the computed third-order elastic constants at
    melting ($\phi=0.54329$). These numbers were obtained for fcc and
    hcp systems containing 13292 particles.
    }
  \label{thirdtable}
\end{table}

The difference between the fcc and hcp elastic constants is surprising
because, already in 1967, Stillinger and
Salsburg~\cite{StillingerElasticity} had pointed out that a simple
free-volume model predicts that the fcc and hcp elastic constants
should be equal. However, they also showed that pair and triplet
correlation effects can lead to differences. Still, we were surprised
by the magnitude of the computed differences, in particular for
$C'_{12}$. To double-check our calculations of the elastic constants,
we performed a second, fully independent calculation where we directly
computed the free energy of the crystals in various states of
deformation. The free energy of the (deformed and undeformed) crystals
was calculated using a 20-point Einstein integration
\cite{FrenkelSmit}. We found that the results obtained by the two
methods were completely consistent.  For example, in Fig.~\ref{figdf},
we show the results of the two calculations for free energy change due
to a deformation of the form
\begin{equation}
  \alpha_{ij}=\left(
  \begin{array}{ccc}
    1+\xi & 0 & 0 \\
    0 & 1/(1+\xi) & 0 \\
    0 & 0 & 1 \\
  \end{array}
\right)
\label{DeformationMatrix}
\end{equation}
To lowest order in $\xi$, $\Delta F/V = (- 2 T_{xx} + C_{11} - C_{12})
\xi^2$, for this deformation.  As the figure shows, the differences in
elastic constants $C'_{11}$ and $C'_{12}$, for fcc and hcp, are so
large that a deformation of 1.2\% is enough to make hcp more stable
than fcc.  The free energy increase of the fcc phase due to a
deformation of 2\% is $\Delta f_{fcc}=1.93(1)\cdot10^{-2}$, while for
hcp it is only $\Delta f_{hcp}=1.66(1)\cdot10^{-2}$.  Fig.~\ref{figdf}
also shows the effect of the third-order elastic constants. To within
the statistical accuracy of our simulations, the relevant third-order
elastic constants (see table \ref{thirdtable}), were found to be the
same for fcc and hcp. Hence, they do not affect the free energy
difference between the two lattices.

\begin{figure}
  \includegraphics[angle=0,width=8.0cm]{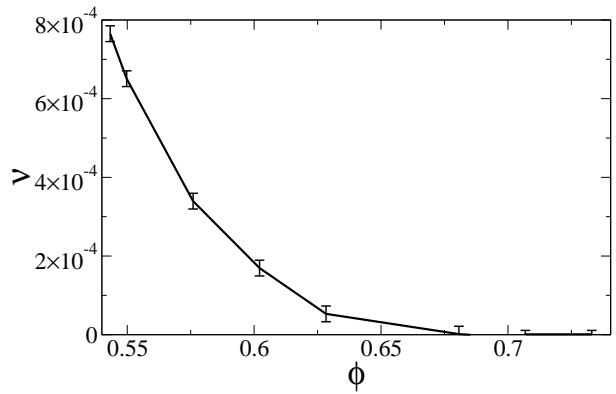}
  \caption{Equilibrium anisotropy ($\nu \equiv 1 -
    \frac{c}{a}/\sqrt{\frac{8}{3}}$)
    for the hcp hard sphere crystal as a function of packing
    fraction.}
  \label{caratio}
\end{figure}

For the undeformed fcc system, all three diagonal components of the
pressure tensor are equal. However, this does not hold for a hcp
system at the same $c/a$-ratio (i.e. for the same spacing between the
close packed [111]-planes).  If we fix the $c/a$ ratio at the fcc
value ($\sqrt{8/3}$), the stresses exhibit a slight anisotropy.  For
the 13292 particle system $T_{xx}$ and $T_{yy}$ are equal (as they
should): $T_{xx}=-11.587(1)$, $T_{yy}=-11.588(1)$. However, $T_{zz}$
is significantly different: $T_{zz}=-11.537(1)$. From
Eq.~\ref{ElasticFreeEnergy}, we can derive what change in the $c/a$
ratio is needed to make the pressure isotropic. We find that, at
melting, isotropy is restored for a $c/a$ ratio of
$\sqrt{8/3}(1-7.5(2)\cdot10^{-4})$.  At higher densities, this value
approaches the close-packing value $c/a=\sqrt{8/3}$, as can be seen in
Fig.~\ref{caratio}. Stillinger and
Salsburg~\cite{StillingerElasticity} used the cell-cluster method to
estimate the difference of the fcc and hcp $c/a$ ratios.  Our
simulations show that, close to melting, the effect is one order of
magnitude larger than predicted.
The free energy difference between the
equilibrium hcp and fcc crystals is only slightly changed by this
relaxation of the hcp $c/a$ ratio: it becomes $1.050(5)\cdot10^{-3}
k_BT$ per particle for $N=13292$ at melting.

As can be seen from the results in table \ref{fulltable} for
$\phi=0.543$ --- where the $c/a$ ratio differs most from fcc --- the
effect of relaxing $c/a$ to its equilibrium value, is barely
significant. For this reason, most hcp elastic constants in table
\ref{fulltable} were computed for $c/a=\sqrt{8/3}$. The table also
shows that the elastic constants depend somewhat on system size, but
the effect is too small to change the qualitative picture.

In colloidal-epitaxy experiments~\cite{JacobHCP}, the best hcp
crystals were obtained when the patterned template was stretched by
$2.6\%$ with respect to the expected lattice spacing at the
experimental packing fraction ($\phi=0.68$).  The templates used
matched a diagonal cut through the $xy$ plane of
Fig.~\ref{coordframe}. Together with the stress produced by gravity
(resulting in a strain perpendicular to the template plane of
$-2.8\%$), this strain is comparable to the strain of
Eq.~\ref{DeformationMatrix} and would result in a free energy
difference of about $3\cdot10^{-2}k_BT$ per particle {\em in favor} of
hcp. The present simulation results may help experimentalists in
designing optimal templates to grow selectively colloidal hcp or fcc
crystals.

We thank Jacob Hoogenboom (Universiteit Twente) for inspiring
discussions about his experimental work.  The work of the FOM
institute is part of the research program of the Foundation for
Fundamental Research on Matter (FOM) and was made possible through
financial support by the Dutch Foundation for Scientific Research
(NWO). 



\end{document}